# Macroscopic State Interferometry Over Large Distances Using State Discrimination

B. T. Kirby and J.D. Franson

*Physics Department, University of Maryland, Baltimore County, Baltimore, MD 21250*

The propagation of macroscopic entangled states over large distances in the presence of loss is of fundamental interest and may have practical applications as well. Here we describe two different techniques in which state discrimination can be used to violate Bell's inequality with macroscopic phase-entangled coherent states. We find that Bell's inequality can be violated by these macroscopic states over a distance of approximately 400 km in commercially-available optical fibers.

**I. Introduction**

The ability to transmit entangled states over large distances would enable a number of important practical applications including quantum key distribution and data transfer between quantum computers. Entanglement is also one of the most fundamental properties of quantum mechanics and tests of nonlocality over large distances are of fundamental interest. This is especially true of experiments involving macroscopic entangled states, which can provide insight into decoherence and the boundary between classical and quantum physics. Here we show that state discrimination techniques [1,2] allow macroscopic entangled states to be transmitted over distances on the order of 400 km in conventional optical fiber while still maintaining sufficient fidelity to violate the Clauser-Horne-Shimony-Holt (CHSH) form of Bell's inequality [3,4].

Our approach is based on the use of weak nonlinearities to generate phase-entangled coherent states [5-11]. A single photon passing through an interferometer can be used to produce a small phase shift in one of two coherent states (laser beams) if a suitable Kerr medium is present in each path through the interferometer. This produces an entangled Schrodinger cat state with anti-correlated phase shifts in the two coherent states. Similar single-photon interferometers located at large distances can then be used to violate the CHSH form of Bell's inequality as described in more detail in the following section. A key feature of this approach is the use of state vector discrimination [1,2] to distinguish between the various phase-shifted coherent states, which greatly increases the range over which Bell's inequality can be violated as compared to a previous approach based on homodyne detection [12].

Decoherence due to photon loss is an important consideration in any approach for transmitting optical entangled states over large distances. Here we model the effects of photon loss by a series of beam splitters, although it can be shown that similar results would be obtained in the presence of absorption by two-level atoms [12,13]. A low rate of decoherence due to photon loss can be achieved for these Schrodinger-cat states when their separation in phase space is relatively small [14].

A macroscopic nonlocal interferometer based on the use of phase-entangled states and homodyne detection is described in Section II. The increased range achievable using a straightforward state vector discrimination technique is analyzed in Section III. A further increase in the range of up to 400 km in optical fiber using a more efficient form of state vector discrimination is then discussed in Section IV. A summary and conclusion are provided in section V.

**II. Nonlocal Interferometry**

In this section we review a nonlocal interferometer based on phase-entangled coherent states and homodyne detection that we previously proposed [12]. The source of the phase-entangled coherent states is illustrated in the left-hand side of Fig. 1 [5-11]. A single photon labelled A passes through a Mach-Zehnder interferometer containing a Kerr medium in each path. A nonlinear phase shift of $2\phi$ is assumed to be generated if the single photon and a coherent state are present in one of the Kerr media simultaneously. A bias phase shift of $-\phi$ is added to both beams so that phase shifts of $\pm\phi$ are created in each beam depending on the path taken by photon A. Post-selection on events in which a photon was observed in detector 1 ensures that there is a well-defined phase between the two terms in the superposition state $|\psi_S\rangle$ that describes the output of the source, which is given by

$$|\psi_S\rangle = (|\alpha_+\rangle|\beta_-\rangle + |\alpha_-\rangle|\beta_+\rangle)/\sqrt{2}. \qquad (1)$$

Here $|\alpha_+\rangle$ represents a coherent state in beam 1 with a positive phase shift while $|\beta_-\rangle$ represents a coherent state in beam 2 with a negative phase shift. The states $|\alpha_-\rangle$ and $|\beta_+\rangle$ are defined in a similar way.

This entangled state can then be probed using two distant interferometers B and C as illustrated in the right-hand side of Fig. 1. Both interferometers have a single Kerr medium placed in one of the two paths, which again produces a phase shift of $2\phi$ if both a coherent state and a single photon are present in the same path. Bias phase shifts of $-\phi$ are added once again so that a net phase shift of $\pm\phi$ is produced depending on the path taken by the single photons as before. In addition, fixed (linear) phase shifts $\sigma_1$ and $\sigma_2$ are included in one of the two paths of each interferometer as shown in the figure.

FIG. 1. A nonlocal interferometer based on the use of weak nonlinearities to produce a small phase shift in a coherent state [12]. The phase shift of the final state can be measured using a homodyne detector as in Ref. [12], while the range over which Bell's inequality can be violated can be greatly increased using state discrimination techniques instead.

Homodyne measurements are then used to determine the final phases of the coherent states after they have passed through both sets of interferometers. Post-selection is performed in which we only accept those events in which detectors 1, 3, and 5 were triggered and in which both coherent states were measured to have a net phase shift of zero. It can be seen that an outcome of that kind can only occur if both photons B and C took the left path or if both of them took the right path. This gives rise to quantum interference between the corresponding probability amplitudes, with a relative phase that depends on the values of $\sigma_1$ and $\sigma_2$. This interference between the left-left and right-right probability amplitudes is analogous to the more familiar long-long and short-short interference that is responsible for the two-photon nonlocal interferometer proposed previously by one of the authors [15].

The state of the system after the photons have passed through the interferometers but before any measurements have been made can be written as

$$|\Psi\rangle = \frac{1}{2^3}\Big[e^{i\sigma_2}|\alpha_{++}\rangle|\beta_{--}\rangle - |\alpha_{++}\rangle|\beta_{-+}\rangle$$
$$-e^{i(\sigma_1+\sigma_2)}|\alpha_{+-}\rangle|\beta_{--}\rangle + e^{i\sigma_1}|\alpha_{+-}\rangle|\beta_{-+}\rangle$$
$$-e^{i\sigma_2}|\alpha_{-+}\rangle|\beta_{+-}\rangle + |\alpha_{-+}\rangle|\beta_{++}\rangle \quad (2)$$
$$+e^{i(\sigma_1+\sigma_2)}|\alpha_{--}\rangle|\beta_{+-}\rangle - e^{i\sigma_1}|\alpha_{--}\rangle|\beta_{++}\rangle\Big]$$
$$\times|1\rangle_1|0\rangle_2|1\rangle_3|0\rangle_4|1\rangle_5|0\rangle_6 + |\psi_\perp\rangle.$$

Here the subscripts on the coherent state amplitudes represent the positive and negative phase shifts produced by the Kerr media and a $\pi/2$ phase shift has been added upon reflection by a beam splitter. The state of the fields in the output ports of the single-photon interferometers are designated by $|1\rangle_i$ if a photon is present in that path and $|0\rangle_i$ if no photons are present (the vacuum state), where $i$ labels the output ports shown in Fig. 1. Only those terms where single photons are present in paths 1, 3, and 5 are explicitly included in Eq. (2), with the remaining orthogonal terms contained in $\psi_\perp$.

If the homodyne measurements are capable of completely distinguishing between these phase-shifted states, then the measurement process can be modeled as a projection onto the states of interest [12]. The corresponding projection for the case in which a photon was detected in detectors 1, 3, and 5 while zero net phase shifts were observed for both coherent states can be written as

$$|p\rangle = \frac{1}{2^3}\Big[e^{i\sigma_1}|\alpha_{+-}\rangle|\beta_{-+}\rangle - e^{i\sigma_2}|\alpha_{-+}\rangle|\beta_{+-}\rangle\Big] \quad (3)$$
$$\times|1\rangle_1|0\rangle_2|1\rangle_3|0\rangle_4|1\rangle_5|0\rangle_6.$$

The probability of such an outcome is given by

$$\langle p|p\rangle = \frac{1}{2^6}\big|e^{i\sigma_1} - e^{i\sigma_2}\big|^2 = \frac{1}{2^4}\sin^2\left(\frac{\sigma_1-\sigma_2}{2}\right). \quad (4)$$

In the absence of any photon loss or measurement noise, this corresponds to an interference pattern with a visibility of 100%, which can be used to violate the CHSH form of the Bell inequality [3,4].

Photon loss reduces the visibility of the interference pattern for two reasons. The first problem is decoherence produced by which-path information left in the environment when a photon is absorbed or scattered out of an optical fiber. The second problem is the increasing overlap of the coherent states as their amplitudes are reduced by loss and they approach the vacuum as illustrated in Fig. (2a). This makes it more difficult to distinguish between the various phase-shifted states.

The effects of photon loss can be included by assuming that beam splitters have been inserted into the long paths between the interferometers. First consider the effects of inserting a single beam splitter with a small reflectivity into the paths to interferometers A and B. If we let $|\gamma_\pm\rangle$ and $|\delta_\pm\rangle$ denote the coherent states in the output ports of the beam splitters in the paths to interferometers A and B respectively, then the projection $|p_L\rangle$ onto the state of interest is given by

$$|p_L\rangle = \frac{1}{2^3}\left[e^{i\sigma_1}|\alpha'_{+-}\rangle|\beta'_{-+}\rangle|\gamma_+\rangle|\delta_-\rangle \\ -e^{i\sigma_2}|\alpha'_{-+}\rangle|\beta'_{+-}\rangle|\gamma_-\rangle|\delta_+\rangle\right]|1\rangle_1|0\rangle_2|1\rangle_3|0\rangle_4|1\rangle_5|0\rangle_6 \quad (5)$$

instead of by Eq. (3). Here the primes in the coherent states $|\alpha'_{+-}\rangle$, $|\alpha'_{-+}\rangle$, $|\beta'_{-+}\rangle$, and $|\beta'_{+-}\rangle$ represent the fact that their amplitudes have been reduced by the beam splitters.

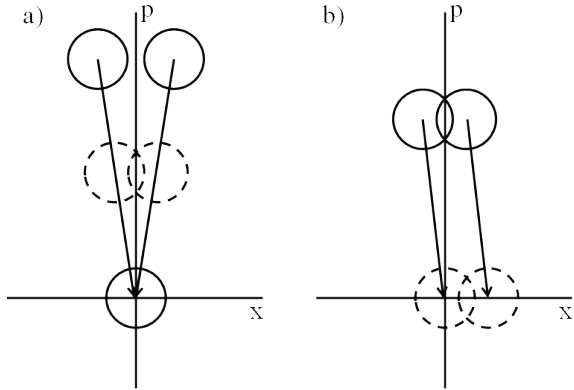

FIG. 2. Phase-space diagrams (Wigner distributions) illustrating two coherent states with different phases. Here x and p represent the two quadratures of the field in dimensionless units. (a) Increased overlap of two coherent states as their amplitude is reduced by photon loss and they approach the vacuum state. (b) Use of displacement operations to transform one of two partially-overlapping coherent states to the vacuum, which allows single-photon detection to distinguish between the two states.

The interference cross terms in $\langle p_L|p_L\rangle$ will be reduced to the extent that there is limited overlap between the states $|\gamma_-\rangle$ and $|\gamma_+\rangle$, for example. As a result, it can be shown that the visibility $v$ of the interference pattern will be reduced to

$$v = |\langle \gamma_+|\gamma_-\rangle|^2 = \exp\left[-|\gamma_+ - \gamma_-|^2\right]. \quad (6)$$

For simplicity, we have assumed that both beams experience the same loss and the square in Eq. (6) reflects the contributions from both beam splitters. We can write $|\gamma_\pm\rangle$ in the form

$$|\gamma_\pm\rangle = |r\alpha e^{\pm i\phi}\rangle, \quad (7)$$

where r is the reflectivity of the beam splitter inserted into the path to interferometer A and $\alpha$ is the initial coherent state amplitude. Then $e^{\pm i\phi}$ terms can be expanded in a Taylor series for small values of $\phi$ which reduces eq. (6) to

$$v = \exp\left[-4(r\alpha\phi)^2\right] = \exp\left[-4N_L\phi^2\right]. \quad (8)$$

Here we have defined $N_L = (r\alpha)^2$ as the average number of photons lost in each path.

This reduction in the visibility can be interpreted as being due to information left in the output ports of the beam splitters. The same results are obtained if a large number of beam splitters produce a total loss of $N_L$ photons in each path.

If homodyne measurements are used to measure the final phase shift of the coherent states, then the increasing overlap of the various phase-shifted states in the presence of loss makes it increasingly difficult to accurately distinguish between them as illustrated in Fig. (2a). This introduces errors into the measured correlations and further reduces the visibility of the interference pattern. That effect was analyzed in detail in Ref. [12], where it was found that the maximum distance over which the CHSH form of Bell's inequality can be violated is limited to roughly 8 km in optical fibers with 0.15 dB/km loss. That analysis will not be described in more detail here because the problem can be avoided by replacing the homodyne measurements with state discrimination techniques as described in the next section.

### III. Unambiguous State Discrimination

A simple example of unambiguous state discrimination is illustrated in Fig. (2b) [1,2]. Here two partially-overlapping coherent states are displaced [16] in such a way that one of them is transformed into the vacuum state. Displacement operations of this kind can be implemented by combining the coherent state of interest with an external laser at a beam splitter in the limit in which the reflectivity of the beam splitter is very small, as illustrated in Fig. 3 [17]. Once one of the coherent states has been displaced to the vacuum in this way, the detection of one or more photons indicates that the other coherent state must have been present. Ignoring the effects of detector noise for the moment, this process allows the two coherent states to be distinguished with certainty some fraction of the time.

Unambiguous state discrimination of this kind has been extensively studied both theoretically [1,2,18-26] and experimentally [27-34]. The usual goal of unambiguous state discrimination is to determine which of two or more phase-shifted coherent states is present.

Our application is somewhat less demanding in the sense that we only need to determine whether or not one particular state was present without necessarily distinguishing between the two remaining coherent states. This results in a success rate that is somewhat higher than would otherwise be the case.

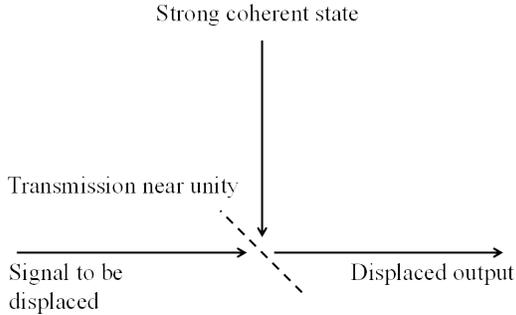

FIG. 3. Implementation of a coherent state displacement operation using a strong reference coherent state (laser beam) combined with the much weaker input coherent state on a beam splitter with a small reflectivity [16,17].

A straightforward state discrimination technique that can be used to post-select those events in which the coherent state from laser 1 has undergone a net phase shift of zero is illustrated in Fig. 4. The interferometers of Fig. 1 will have produced a net phase shift of either $\pm 2\phi$ or 0 depending on the paths taken by the single photons. The coherent state at the output of interferometer B is first passed through a 50/50 beam splitter. A displacement operation is then performed on the coherent state in one of the output ports of the beam splitter in such a way as to displace a state with phase shift $2\phi$ to the vacuum. The detection of one or more photons after that displacement operation indicates that a state with phase shift $2\phi$ was not present. The coherent state in the other output port of the beam splitter is then displaced in such a way that a state with phase shift $-2\phi$ will be displaced to the vacuum, and the detection of one or more photons there indicates that a state with that phase was not present. We post-select on those events where one or more photons were detected in both output ports of the beam splitter, which can only occur when the coherent state had zero net phase shift as desired.

A similar state discrimination technique is also applied to the coherent state from laser 2. A successful outcome for both measurements can then be used to post-select the two states shown in Eq. (3) as required for quantum interference to occur. This requires two successful detection events for the coherent state from laser 1 and two more detection events for the coherent state from laser 2. This dependence on four-fold detection events gives a relatively low success rate when both signals have been highly attenuated by the losses in an optical fiber. A more efficient state discrimination technique will be described in the following section, but we will first analyze the straightforward approach described above.

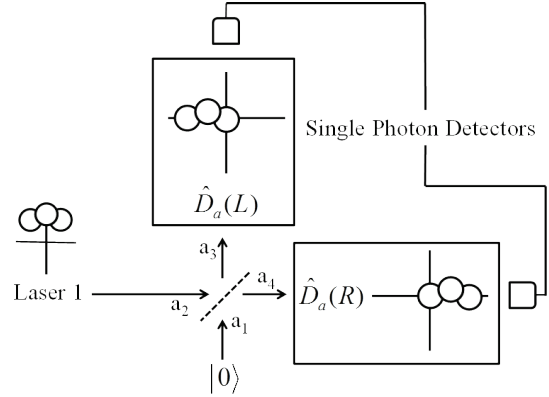

FIG. 4. Use of state discrimination techniques to determine whether or not the coherent state from laser 1 has undergone a net phase shift of zero. A 50/50 beam splitter divides the signal into two identical coherent states. Displacement operations performed on the coherent states in the two output ports of the beam splitter are then used to rule out states with phase shifts of $\pm\phi$. If one or more photons are detected in both output ports, then the coherent state must have had a net phase shift of zero as desired.

Let operator $\hat{B}(\lambda)$ denote the effect of a beam splitter with reflectivity $\lambda$ acting on two incident coherent states $|\mu\rangle$ and $|\nu\rangle$ in input ports 1 and 2, respectively [26]:

$$\hat{B}(\lambda)|\mu\rangle_1|\nu\rangle_2 = \left|\sqrt{1-\lambda}\mu + \sqrt{\lambda}\nu\right\rangle_3 \otimes \left|-\sqrt{\lambda}\mu + \sqrt{1-\lambda}\nu\right\rangle_4. \quad (9)$$

For the case of a vacuum state in port 1 and a 50/50 beam splitter ($\lambda = 1/2$), this simplifies to

$$\hat{B}(1/2)|0\rangle_1|\nu\rangle_2 = \left|\frac{1}{\sqrt{2}}\nu\right\rangle_3 \otimes \left|\frac{1}{\sqrt{2}}\nu\right\rangle_4. \quad (10)$$

A displacement operator $\hat{D}(\tau)$ acting on a coherent state $|\nu\rangle$ is defined by [16]

$$\hat{D}(\tau)|\nu\rangle = |\nu + \tau\rangle, \quad (11)$$

where both $\nu$ and $\tau$ are in general complex numbers.

The state of the system before the measurements shown in Fig. 4 can be written as in Eq. (2), but including the loss terms $|\gamma_\pm\rangle$ and $|\delta_\pm\rangle$ discussed in section II gives

$$|\Psi'\rangle = \frac{1}{2^3}\Big[ e^{i\sigma_2}|\alpha'_{++}\rangle|\beta'_{--}\rangle|\gamma_+\rangle|\delta_-\rangle - |\alpha'_{++}\rangle|\beta'_{-+}\rangle|\gamma_+\rangle|\delta_-\rangle$$
$$-e^{i(\sigma_1+\sigma_2)}|\alpha'_{+-}\rangle|\beta'_{--}\rangle|\gamma_+\rangle|\delta_-\rangle + e^{i\sigma_1}|\alpha'_{+-}\rangle|\beta'_{-+}\rangle|\gamma_+\rangle|\delta_-\rangle$$
$$-e^{i\sigma_2}|\alpha'_{-+}\rangle|\beta'_{+-}\rangle|\gamma_-\rangle|\delta_+\rangle + |\alpha'_{-+}\rangle|\beta'_{++}\rangle|\gamma_-\rangle|\delta_+\rangle$$
$$+e^{i(\sigma_1+\sigma_2)}|\alpha'_{--}\rangle|\beta'_{+-}\rangle|\gamma_-\rangle|\delta_+\rangle - e^{i\sigma_1}|\alpha'_{--}\rangle|\beta'_{++}\rangle|\gamma_-\rangle|\delta_+\rangle \Big].$$
(12)

Here the single photon and orthogonal terms have been dropped for convenience.

The amplitudes $\tilde{\alpha}_{\pm\pm}$ of the coherent states in the output ports of the beam splitter shown in Fig. 4 will correspond to the amplitudes of Eq. (12) reduced by a factor of $1/\sqrt{2}$. Using $\exp(2i\phi) = \cos(2\phi) + i\sin(2\phi)$ gives

$$\tilde{\alpha}_{+-} = \tilde{\alpha}_{-+} = i\frac{|\alpha'|}{\sqrt{2}}$$
$$\tilde{\alpha}_{++} = -\frac{|\alpha'|}{\sqrt{2}}\sin(2\phi) + i\frac{|\alpha'|}{\sqrt{2}}\cos(2\phi) \quad (13)$$
$$\tilde{\alpha}_{--} = \frac{|\alpha'|}{\sqrt{2}}\sin(2\phi) + i\frac{|\alpha'|}{\sqrt{2}}\cos(2\phi).$$

Here we have defined $\alpha' \equiv \alpha'_{+-} = \alpha'_{-+}$. Similar results apply to the coherent state from laser 2.

We will denote the two displacement operations shown in Fig. 4 by $\hat{D}(L)$ and $\hat{D}(R)$. The desired effects of these displacement operations on the coherent states from beam 1 will be denoted as follows:

$$\hat{D}(L)|\tilde{\alpha}_{++}\rangle = |L'_+\rangle \qquad \hat{D}(R)|\tilde{\alpha}_{++}\rangle = |0\rangle$$
$$\hat{D}(L)|\tilde{\alpha}_{\pm\mp}\rangle = |L'_0\rangle \qquad \hat{D}(R)|\tilde{\alpha}_{\pm\mp}\rangle = |R'_0\rangle \quad (14)$$
$$\hat{D}(L)|\tilde{\alpha}_{--}\rangle = |0\rangle \qquad \hat{D}(R)|\tilde{\alpha}_{--}\rangle = |R'_-\rangle.$$

Here we have used $|L_+'\rangle$ to denote the state of the positively phase-shifted state after the displacement operation, with a similar notation for the other states.

Combining Eqs. (11), (13), and (14) gives the required values of the displacement amplitudes L and R:

$$L = -\frac{|\alpha'|}{\sqrt{2}}\sin(2\phi) - i\frac{|\alpha'|}{\sqrt{2}}\cos(2\phi)$$
$$R = \frac{|\alpha'|}{\sqrt{2}}\sin(2\phi) - i\frac{|\alpha'|}{\sqrt{2}}\cos(2\phi).$$
(15)

The $L_0'$ and $R_0'$ amplitudes will play an essential role in what follows. Applying Eq. (15) to the amplitudes of Eq. (13) gives their values as

$$L_0' = -\frac{|\alpha'|}{\sqrt{2}}\sin(2\phi) + i\frac{|\alpha'|}{\sqrt{2}}[1-\cos(2\phi)]$$
$$R_0' = \frac{|\alpha'|}{\sqrt{2}}\sin(2\phi) + i\frac{|\alpha'|}{\sqrt{2}}[1-\cos(2\phi)].$$
(16)

The state of the system just before the single photon detectors can be found by applying the relevant beam splitter and displacement operators given above to the state of the system in Eq. (12). The beam splitter operators for beams 1 and 2 will be denoted $\hat{B}_a(1/2)$ and $\hat{B}_b(1/2)$ respectively, where the subscripts $a$ and $b$ refer to the output of interferometers A and B. The combined result of all of the beam splitter and displacement operations is then given by

$$\hat{D}_{a_3b_3}(L)\hat{D}_{a_4b_4}(R)\hat{B}_a(1/2)\hat{B}_b(1/2)|\Psi'\rangle =$$
$$\frac{1}{2^3}\Big[ e^{i\sigma_2}\big(|L_+'\rangle_{a_3}\otimes|0\rangle_{a_4}\big)\big(|0\rangle_{b_3}\otimes|R_-'\rangle_{b_4}\big)|\gamma_+\rangle|\delta_-\rangle$$
$$-\big(|L_+'\rangle_{a_3}\otimes|0\rangle_{a_4}\big)\big(|L_0'\rangle_{b_3}\otimes|R_0'\rangle_{b_4}\big)|\gamma_+\rangle|\delta_-\rangle$$
$$-e^{i(\sigma_1+\sigma_2)}\big(|L_0'\rangle_{a_3}\otimes|R_0'\rangle_{a_4}\big)\big(|0\rangle_{b_3}\otimes|R_-'\rangle_{b_4}\big)|\gamma_+\rangle|\delta_-\rangle$$
$$+e^{i\sigma_1}\big(|L_0'\rangle_{a_3}\otimes|R_0'\rangle_{a_4}\big)\big(|L_0'\rangle_{b_3}\otimes|R_0'\rangle_{b_4}\big)|\gamma_+\rangle|\delta_-\rangle \quad (17)$$
$$-e^{i\sigma_2}\big(|L_0'\rangle_{a_3}\otimes|R_0'\rangle_{a_4}\big)\big(|L_0'\rangle_{b_3}\otimes|R_0'\rangle_{b_4}\big)|\gamma_-\rangle|\delta_+\rangle$$
$$+\big(|L_0'\rangle_{a_3}\otimes|R_0'\rangle_{a_4}\big)\big(|L_+'\rangle_{b_3}\otimes|0\rangle_{b_4}\big)|\gamma_-\rangle|\delta_+\rangle$$
$$+e^{i(\sigma_1+\sigma_2)}\big(|0\rangle_{a_3}\otimes|R_-'\rangle_{a_4}\big)\big(|L_0'\rangle_{b_3}\otimes|R_0'\rangle_{b_4}\big)|\gamma_-\rangle|\delta_+\rangle$$
$$-e^{i\sigma_1}\big(|0\rangle_{a_3}\otimes|R_-'\rangle_{a_4}\big)\big(|L_+'\rangle_{b_3}\otimes|0\rangle_{b_4}\big)|\gamma_-\rangle|\delta_+\rangle \Big],$$

where it was assumed that $\alpha = \beta$ for convenience.

We will consider the case in which the coherent states have been attenuated to the point that there is a negligible probability of detecting more than one photon in any of the single-photon detectors shown in Fig. 4. The projection of Eq. (17) onto a state in which there is a single photon in each of the four detectors gives

$$\langle 1,1,1,1|\hat{D}_{a_3b_3}(L)\hat{D}_{a_4b_4}(R)\hat{B}_a(1/2)\hat{B}_b(1/2)|\Psi'\rangle =$$
$$\frac{1}{2^3}\Big[ e^{i\sigma_1}\langle 1|L_0'\rangle\langle 1|R_0'\rangle\langle 1|L_0'\rangle\langle 1|R_0'\rangle|\gamma_+\rangle|\delta_-\rangle$$
$$-e^{i\sigma_2}\langle 1|L_0'\rangle\langle 1|R_0'\rangle\langle 1|L_0'\rangle\langle 1|R_0'\rangle|\gamma_-\rangle|\delta_+\rangle \Big].$$
(18)

Here the state $|1,1,1,1\rangle$ corresponds to having a single photon in each of the detectors while the state $|1\rangle$ denotes the presence of a photon in the individual detectors.

Factoring out the common terms reduces Eq. (18) to

$$\langle 1,1,1,1|\hat{D}_{a_3,b_3}(L)\hat{D}_{a_4,b_4}(R)\hat{B}_a\left(\frac{1}{2}\right)\hat{B}_b\left(\frac{1}{2}\right)|\Psi'\rangle = \frac{(\langle 1|L'_0\rangle\langle 1|R'_0\rangle)^2}{2^3}\left[e^{i\sigma_1}|\gamma_+\rangle|\delta_-\rangle - e^{i\sigma_2}|\gamma_-\rangle|\delta_+\rangle\right] \quad (19)$$

which is similar in form to Eqs. (3) and (5). The probability $P_s$ of a successful detection event is given by

$$P_s = \left|\langle 1,1,1,1|\hat{D}_{a_3,b_3}(L)\hat{D}_{a_4,b_4}(R)\hat{B}_a\left(\frac{1}{2}\right)\hat{B}_b\left(\frac{1}{2}\right)|\Psi'\rangle\right|^2$$
$$= \frac{|\langle 1|L'_0\rangle\langle 1|R'_0\rangle|^4}{2^5}\left[1 - |\langle\gamma_+|\gamma_-\rangle|^2\cos(\sigma_1-\sigma_2)\right], \quad (20)$$

where it was assumed once again that the same loss is experienced by both beams ($|\gamma|=|\delta|$). This corresponds to a visibility of $v=|\langle\gamma_+|\gamma_-\rangle|^2=\exp[-4N_L\phi^2]$ which is the same as that in Eq. (8). The factors of $\langle 1|L_0'\rangle$ and $\langle 1|R_0'\rangle$ only affect the counting rate and not the visibility. This represents a major advantage over the use of homodyne measurements, where the overlap of the coherent states in the presence of loss produces a further decrease in the visibility.

The $\langle 1|L_0'\rangle$ and $\langle 1|R_0'\rangle$ factors in Eq. (20) can be evaluated using Eq. (16), which gives

$$\langle 1|L'_0\rangle = \frac{|\alpha'|}{\sqrt{2}}\left[-\sin(2\phi)+i(1-\cos(2\phi))\right]e^{-|\alpha'|^2\sin^2(\phi)}$$
$$\langle 1|R'_0\rangle = \frac{|\alpha'|}{\sqrt{2}}\left[\sin(2\phi)+i(1-\cos(2\phi))\right]e^{-|\alpha'|^2\sin^2(\phi)}. \quad (21)$$

Inserting this into Eq. (20) gives

$$P_s = \frac{|\alpha'|^8\sin^8(\phi)e^{-8|\alpha'|^2\sin^2(\phi)}}{2}\left[1-e^{-4N_L\phi^2}\cos(\sigma_1-\sigma_2)\right]. \quad (22)$$

As an example, consider the case in which $\alpha=100$, $\phi=0.0028$, there is a loss of 0.15 dB/km in the optical fibers, and a total distance of 140 km between interferometers B and C (70 km from the source to each interferometer). Then $|\alpha'|$ can be found from the relation $|\alpha'|^2=|\alpha|^2 10^{-0.15*70/10}=891.251$. After the coherent states in each arm have travelled 70 km the number of photons lost in each of the beams is given by $|\alpha|^2-|\alpha'|^2=9108.75=N_L$. Inserting these values into Eq. (22) with $\sigma_1$ and $\sigma_2$ chosen to give the maximum $R_{max}$ or minimum $R_{min}$ counting rates gives

$$R_{max} = 1.97\times 10^{-9} \quad (\sigma_1-\sigma_2=\pi)$$
$$R_{min} = 0.28\times 10^{-9} \quad (\sigma_1-\sigma_2=0). \quad (23)$$

Assuming a source that operates at a rate of 1 GHz, we can expect approximately 2 coincidence counts per second at the maximum of the interference pattern and 0.3 counts per second when at the minimum. This corresponds to a visibility of 75%, which is in agreement with Eq. (8) and above the 70.7% value needed to violate the CHSH form of Bell's inequality [3,4].

**IV. Enhanced Approach**

The state discrimination approach described above has the advantage that the visibility of the interference pattern is not affected by the increased overlap between the phase-shifted coherent states due to photon loss, but the success rate is relatively low due to its dependence on the detection of a total of four photons from the displaced coherent states. Here we describe an enhanced approach that only requires the detection of two photons in the displaced coherent states, which substantially increases the useful range of the system.

The enhanced state discrimination approach is illustrated in Fig. 5. As before, each of the two coherent states will have been shifted in phase by $\pm 2\phi$ or 0 and we need to be able to distinguish between the various phase-shifted states. Here each of the coherent states is displaced in such a way that the states with zero net phase shift are displaced to the vacuum. No additional beam splitters of the kind shown in Fig. 4 are required. A detection of one or more photons in the displaced coherent states from both laser 1 and laser 2 eliminates the probability amplitude for the terms in Eq. (12) that correspond to zero net phase shift. The only two terms that remain in the post-selected state now involve $|\alpha_{++}\rangle|\beta_{--}\rangle$ and $|\alpha_{--}\rangle|\beta_{++}\rangle$, whereas the original approach involved $|\alpha_{+-}\rangle|\beta_{-+}\rangle$ and $|\alpha_{-+}\rangle|\beta_{+-}\rangle$ instead. Quantum interference between the corresponding probability amplitudes can once again violate the CSHS form of Bell's inequality as described in more detail below.

The amplitudes of the three possible coherent states from laser 1 before the displacement operations are a factor of $\sqrt{2}$ larger than those given by Eq. (13) due to the absence of the beam splitter in Fig. 4 in this enhanced approach. It can be seen that the displacement operation needed to transform the state with zero net phase shift into the vacuum state is given by $\hat{D}(-i|\alpha'|)$. The effect of this displacement on the states produced by laser 1 is then

$$\hat{D}(-i|\alpha'|)|\alpha'\rangle = |0\rangle$$
$$\hat{D}(-i|\alpha'|)|\alpha'_{--}\rangle = ||\alpha'|\sin(2\phi) + i|\alpha'|(\cos(2\phi)-1)\rangle = |\alpha'_{D-}\rangle$$
$$\hat{D}(-i|\alpha'|)|\alpha'_{++}\rangle = |-|\alpha'|\sin(2\phi) + i|\alpha'|(\cos(2\phi)-1)\rangle = |\alpha'_{D+}\rangle$$
(24)

with similar results for beam 2.

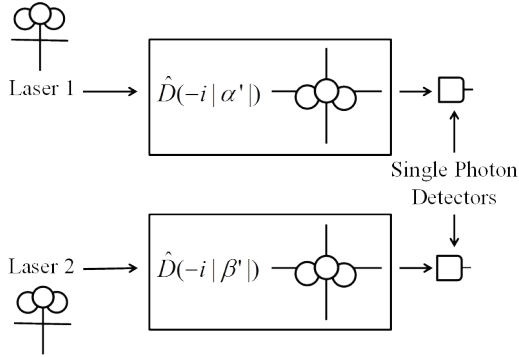

FIG. 5. Enhanced state discrimination technique in which the coherent states from each of the laser beams are displaced in such a way that the states with zero net phase shift are transformed into the vacuum state. The detection of one or more photons from both of the displaced coherent states rules out the possibility of all but two terms in the post-selected state, and those terms can then interfere to violate Bell's inequality.

Applying these displacement operators to both beams 1 and 2 in Eq. (12) results in

$$\hat{D}_1(-i|\alpha'|)\hat{D}_2(-i|\beta'|)|\Psi'\rangle =$$
$$\frac{1}{2^3}\Big[e^{i\sigma_2}|\alpha'_{D+}\rangle|\beta'_{D-}\rangle|\gamma_+\rangle|\delta_-\rangle$$
$$-|\alpha'_{D+}\rangle|0\rangle|\gamma_+\rangle|\delta_-\rangle$$
$$-e^{i(\sigma_1+\sigma_2)}|0\rangle|\beta'_{D-}\rangle|\gamma_+\rangle|\delta_-\rangle$$
$$+e^{i\sigma_1}|0\rangle|0\rangle|\gamma_+\rangle|\delta_-\rangle$$
$$-e^{i\sigma_2}|0\rangle|0\rangle|\gamma_-\rangle|\delta_+\rangle$$
$$+|0\rangle|\beta'_{D+}\rangle|\gamma_-\rangle|\delta_+\rangle$$
$$+e^{i(\sigma_1+\sigma_2)}|\alpha'_{D-}\rangle|0\rangle|\gamma_-\rangle|\delta_+\rangle$$
$$-e^{i\sigma_1}|\alpha'_{D-}\rangle|\beta'_{D+}\rangle|\gamma_-\rangle|\delta_+\rangle\Big].$$
(25)

Here we have defined the displaced states $|\alpha_{D\pm}\rangle$ and $|\beta_{D\pm}\rangle$ as indicated in Eq. (24). The probability of detecting a single photon in both beam 1 and beam 2 after the displacements of Fig. 5 can be found by projecting Eq. (25) onto single photon states, where we have assumed once again that the coherent states are sufficiently weak that we can neglect the probability of there being more than one photon in either detector. This gives

$$\langle 1,1|\hat{D}_1(-i|\alpha'|)\hat{D}_2(-i|\beta'|)|\Psi'\rangle =$$
$$\frac{1}{2^3}\Big[e^{i\sigma_2}\langle 1|\alpha'_{D+}\rangle\langle 1|\beta'_{D-}\rangle|\gamma_+\rangle|\delta_-\rangle$$
$$-e^{i\sigma_1}\langle 1|\alpha'_{D-}\rangle\langle 1|\beta'_{D+}\rangle|\gamma_-\rangle|\delta_+\rangle\Big],$$
(26)

where the notation is analogous to that in Eq. (18).

The detection probability $P_D$ is then given by

$$P_D = |\langle 1,1|\hat{D}_1(-i|\alpha'|)\hat{D}_2(-i|\beta'|)|\Psi'\rangle|^2$$
$$= \frac{1}{2^6}\Big[\langle\alpha'_{D+}|1\rangle\langle 1|\alpha'_{D+}\rangle\langle\beta'_{D-}|1\rangle\langle 1|\beta'_{D-}\rangle$$
$$+\langle\alpha'_{D-}|1\rangle\langle 1|\alpha'_{D-}\rangle\langle\beta'_{D+}|1\rangle\langle 1|\beta'_{D+}\rangle$$
$$-e^{-i(\sigma_1-\sigma_2)}\langle 1|\alpha'_{D+}\rangle\langle 1|\beta'_{D-}\rangle\langle\alpha'_{D-}|1\rangle\langle\beta'_{D+}|1\rangle\langle\gamma_-|\gamma_+\rangle\langle\delta_+|\delta_-\rangle$$
$$-e^{i(\sigma_1-\sigma_2)}\langle\alpha'_{D+}|1\rangle\langle\beta'_{D-}|1\rangle\langle 1|\alpha'_{D-}\rangle\langle 1|\beta'_{D+}\rangle\langle\gamma_+|\gamma_-\rangle\langle\delta_-|\delta_+\rangle\Big].$$
(27)

Assuming once again that both lasers have the same initial amplitude ($\alpha = \beta$) and experience the same loss ($\gamma = \delta$), this reduces to

$$P_D = \frac{|\langle 1|\alpha'_{D+}\rangle|^2 |\langle 1|\alpha'_{D-}\rangle|^2}{2^5}\Big[1-|\langle\gamma_-|\gamma_+\rangle|^2\cos(\sigma_1-\sigma_2)\Big] \quad (28)$$

The amplitudes $\alpha'_{D+}$ and $\alpha'_{D-}$ are displaced by equal amounts so that $|\langle 1|\alpha'_{D+}\rangle|^2 = |\langle 1|\alpha'_{D-}\rangle|^2$. The single-photon term in the usual expression for a coherent states gives

$$|\langle 1|\alpha'_{D-}\rangle|^2 = 4|\alpha'|^2 \sin^2(\phi) e^{-4|\alpha'|^2 \sin^2(\phi)}. \quad (29)$$

This can be inserted into Eq. (28) to give

$$P_D = \frac{|\langle 1|\alpha'_{D-}\rangle|^4}{2^5}[1-|\langle\gamma_-|\gamma_+\rangle|^2\cos(\sigma_1-\sigma_2)]$$
$$= \frac{|\alpha'|^4 \sin^4(\phi) e^{-8|\alpha'|^2 \sin^2(\phi)}}{2}[1-e^{-4N_L\phi^2}\cos(\sigma_1-\sigma_2)].$$
(30)

It can be seen that the visibility of the interference pattern from this approach is the same as that from the previous approach as given in Eq. (8). But the success rate is proportional to $|\alpha'\phi|^4$ rather than $|\alpha'\phi|^8$, which is a considerable improvement given that $|\alpha'\phi|$ is typically much less than 1. As an example, consider a situation in

which $\alpha = 100$, $\phi = 0.0028$, a loss of 0.15 dB/km in optical fiber, and a total separation of 400 km between interferometers B and C. Then $|\alpha'|$ can be found using $|\alpha'|^2 = |\alpha|^2 \, 10^{-0.15*200/10} = 10$. After the coherent states in each path have propagated 200 km, the number of photons lost in each beam is $|\alpha|^2 - |\alpha'|^2 = 9990 = N_L$. Inserting these values into Eq. (30) gives maximum and minimum coincidence rates of

$$R_{\max} = 5.3 \times 10^{-9} \left( \sigma_1 - \sigma_2 = \pi \right)$$
$$R_{\min} = 0.83 \times 10^{-9} \left( \sigma_1 - \sigma_2 = 0 \right). \quad (31)$$

A source operating at a rate of 1 GHz would thus produce 5.3 coincidence counts per second when the phase shifts $\sigma_1$ and $\sigma_2$ are set to give a maximum and 0.8 counts per second when set to give a minimum. This corresponds to a visibility of $v = \exp[-4N_L \phi^2] = 73\%$, which is above the 70.7% value needed to violate the CHSH form of Bell's inequality [3,4].

This enhanced state discrimination technique essentially doubles the range over which the same coincidence counting rate can be obtained as compared to the previous approach described in Section III. In both cases, the range over which Bell's inequality can be violated in is limited only by the desired coincidence rate, which must be sufficiently large compared to the accidental rate in the detectors. The accidental coincidence counting rate due to dark counts is negligible for most single-photon detectors compared to the rates expected from the example considered above. Detector dark counts as low as 0.0008 counts/s [35] have been observed in silicon avalanche photodiodes, for example, with an even lower rate of accidental coincidences.

## V. Summary and Discussion

We have described two ways in which quantum state discrimination can be used to violate the CHSH form of Bell's inequality over large distances using macroscopic phase-entangled coherent states [5-11]. A single photon in an interferometer containing a small Kerr nonlinearity can produce anti-correlated phase shifts in two coherent states. The entanglement between the two coherent states can be probed using two separated single-photon interferometers containing additional Kerr media. Bell's inequality can then be violated by using state discrimination techniques [1,2] to post-select those events in which the coherent states had a specific net phase shift.

The most straightforward state discrimination approach displaces the coherent states in such a way that we can post-select on events in which there was zero net phase shift. This produces quantum interference between the probability amplitudes for the two ways in which that may have occurred. The use of state discrimination techniques in this way greatly increases the range over which Bell's inequality can be violated as compared to the use of homodyne detection, but it does require the detection of a total of four photons from the displaced coherent states.

A more efficient state discrimination technique is based on post-selection of those events in which there was a nonzero phase shift in both coherent states. This only requires the detection of two photons from the displaced coherent states, which increases the rate of success. Once again, quantum interference between the ways in which this can occur allows violations of Bell's inequality. Using this approach, Bell's inequality can be violated over a distance of approximately 400 km in optical fibers using macroscopic phase-entangled states. As is the case for the nonlocal interferometer previously proposed by one of the authors [15], this approach is relatively insensitive to polarization changes during propagation through the optical fibers.

The observation of nonlocal macroscopic quantum effects is of fundamental importance. The approach described here allows the nonlocal nature of entangled Schrodinger cat states to be observed over large distances as a violation of the CHSH form of Bell's inequality. Experiments of this kind may provide additional information regarding possible decoherence mechanisms for entangled macroscopic states propagating over large distances.

These techniques may also have applications in quantum key distribution and quantum communications. The security of quantum key distribution based on this approach will be described in a subsequent paper.

## Acknowledgements

We would like to acknowledge valuable discussions with F. E. Becerra, C. J. Broadbent, G. T. Hickman, D. E. Jones, and T.B. Pittman. This work was supported in part by DARPA DSO under Grant No. W31P4Q-10-1-0018.

## References

[1] S. Dolinar, Research Laboratory of Electronics, MIT, Quarterly Progress Report No. **111**, 1973 (unpublished), p. 115.
[2] R. S. Kennedy, Research Laboratory of Electronics, MIT, Quarterly Progress Report No. **108**, 1973 (unpublished), p. 219.
[3] J. F. Clauser, M. A. Horne, A. Shimony, and R. A. Holt, Phys. Rev. Lett. **23**, 880 (1969).
[4] A. Mann, B. C. Sanders, and W. J. Munro, Phys. Rev. A **51**, 989 (1995).
[5] B. C. Sanders, Phys. Rev. A **45**, 6811 (1992).

[6] C. C. Gerry, Phys. Rev. A **59**, 4095 (1999); C. C. Gerry and R. Grobe, ibid. **75**, 034303 (2007).
[7] H. Jeong, M. S. Kim, T. C. Ralph, and B. S. Ham, Phys. Rev. A **70**, 061801 (2004).
[8] K. Nemoto and W. J. Munro, Phys. Rev. Lett. **93**, 250502 (2004).
[9] W. J. Munro, K. Nemoto, and T. P. Spiller, New J. Phys. **7**, 137 (2005).
[10] H. Jeong, Phys. Rev. A **72**, 034305 (2005).
[11] D. S. Simon, G. Jaeger, and A. V. Sergienko, Phys. Rev. A. **89**, 012315 (2014).
[12] B. T. Kirby and J. D. Franson, Phys. Rev. A **87**, 053822 (2013).
[13] J. D. Franson, Phys. Rev. A **84**, 043831 (2011).
[14] W. Schleich, M. Pernigo, and F. Le Kien, Phys. Rev. A **44**, 2172 (1991).
[15] J. D. Franson, Phys. Rev. Lett. **62**, 2205 (1989).
[16] M. O. Scully and M. S. Zubairy, Quantum Optics (Cambridge University Press, Cambridge, England, 2006).
[17] M. G. A. Paris, Phys. Lett. A **217**, 78 (1996).
[18] I. D. Ivanovic, Phys. Lett. A **123**, 257 (1987).
[19] D. Dieks, Phys. Lett. A **126**, 303 (1988).
[20] A. Peres, Phys. Lett. A **128**, 19 (1988).
[21] K. Banaszek, Phys. Lett. A **253**, 12 (1999).
[22] S. J. van Enk, Phys. Rev. A **66**, 042313 (2002).
[23] H. Jeong and M. S. Kim, Phys. Rev. A **65**, 042305 (2002).
[24] P. van Loock, N. Lutkenhaus, W. J. Munro, and K. Nemoto, Phys. Rev. A **78**, 062319 (2008).
[25] C. Wittmann, U. L. Anderson, and G. Leuchs, J. Mod. Opt. **57**, 213 (2010).
[26] S. Izumi, M. Takeoka, M. Fujiwara, N. Dalla Pozza, A. Assalini, K. Ema, and M. Sasaki, Phys. Rev. A **86**, 042328 (2012).
[27] R. L. Cook, P. J. Martin, and J. M. Geremia, Nature (London) **446**, 774 (2007).
[28] C. Wittmann, M. Takeoka, K. N. Cassemiro, M. Sasaki, G. Leuchs, and U. L. Andersen, Phys. Rev. Lett. **101**, 210501 (2008).
[29] K. Tsujino, D. Fukuda, G. Fujii, S. Inoue, M. Fujiwara, M. Takeoka, and M. Sasaki, Opt. Express **18**, 8107 (2010).
[30] C. Wittmann, U. L. Andersen, M. Takeoka, D. Sych, and G. Leuchs, Phys. Rev. Lett. **104**, 100505 (2010).
[31] F. E. Becerra, J. Fan, G. Baumgartner, S. V. Polyakov, J. Goldhar, J. T. Kosloski, and A. Migdall. Phys. Rev. A **84**, 062324 (2011).
[32] K. Tsujino, D. Fukuda, G. Fujii, S. Inoue, M. Fujiwara, M. Takeoka, and M. Sasaki, Phys. Rev. Lett. **106**, 250503 (2011).
[33] C. R. Muller, M. A. Usuga, C. Wittmann, M. Takeoka, Ch. Marquardt, U. L. Andersen, and G. Leuchs, New J. Phys. **14**, 083009 (2012).
[34] F. E. Becerra, J. Fan, G. Baumgartner, J. Goldhar, J. T. Kosloski, and A. Migdall, Naure Photon. **7**, 147 (2013).
[35] M. Akiba, K. Tsujino, and M. Sasaki, Opt. Lett. **35**, 2621 (2010).